\documentclass[english,twocolumn,transaction]{IEEEtran}
\usepackage[T1]{fontenc}
\usepackage{float}
\usepackage{amsmath}
\usepackage{amssymb}
\usepackage{graphicx}
\usepackage{url}
\usepackage{color}
\usepackage{multirow}

\makeatletter

\floatstyle{ruled}
\newfloat{algorithm}{tbp}{loa}
\providecommand{\algorithmname}{Algorithm}
\floatname{algorithm}{\protect\algorithmname}

\IEEEoverridecommandlockouts
\usepackage{ifpdf}
\usepackage{cite}
\ifCLASSOPTIONcompsoc
\usepackage[caption=false,font=normalsize,labelfont=sf,textfont=sf]{subfig}
\else
\usepackage[caption=false,font=footnotesize]{subfig}
\fi
\usepackage{amsmath}
\makeatother
\usepackage{caption}
\usepackage{babel}
\usepackage{flushend}

\begin{document}
\title{Caching Scalable Videos in the Edge of Wireless Cellular Networks}
\author{Xuewei~Zhang,~Yuan~Ren,~Tiejun~Lv,~\IEEEmembership{Senior~Member,~IEEE}, ~and~Lajos~Hanzo,~\IEEEmembership{Fellow,~IEEE}
\thanks{
Manuscript received August 10, 2021; revised May 7, 2022; accepted July 27, 2022. 
\emph{(Corresponding author: Tiejun Lv and Lajos Hanzo)}.
	
X. Zhang and Y. Ren are with the School of Communications and Information
Engineering, Xi'an University of Posts and Telecommunications (XUPT), Xi'an,
China (e-mail: \{zhangxw,renyuan\}@xupt.edu.cn).

T. Lv is with the School of Information and Communication
Engineering, Beijing University of Posts and Telecommunications (BUPT), Beijing,
China (e-mail: lvtiejun@bupt.edu.cn).

L. Hanzo is with
University of Southampton, Southampton, U.K. (email: lh@ecs.soton.ac.uk).
}}
\maketitle
\begin{abstract}
By pre-fetching popular videos into the local caches of edge nodes,
wireless edge caching provides an effective means of reducing repeated content deliveries.
To meet the various viewing quality requirements of multimedia users,
scalable video coding (SVC) is integrated with edge caching,
where the constituent layers of scalable videos are flexibly cached and transmitted to users.
In this article, we discuss the challenges
arising from the different content popularity and various viewing requirements
of scalable videos,
and present the diverse types of cached contents as well as the corresponding transmission schemes.
We provide an overview of the existing caching schemes,
and summarize the criteria of making caching decisions.
A case study is then presented,
where the transmission delay is quantified and used as the performance metric.
Simulation results confirm that
giving cognizance to the realistic requirements of end users is capable of
significantly reducing the content transmission delay,
compared to the existing caching schemes operating without SVC.
The results also verify that
the transmission delay of the proposed random caching scheme is lower than
that of the caching scheme which only provides local caching gain.
\end{abstract}
\section{Introduction}
\renewcommand\figurename{Fig.}
The world is witnessing a brand-new information age,
where data traffic is escalating rapidly right across the globe \cite{Fan2020Cache}.
The pressure on the network backhaul
which delivers data traffic from the core network to wireless edge nodes
is also escalating,
resulting in the degradation of the quality of service (QoS) provided to end users.
Multimedia video services account for the majority of the increased global data traffic,
and the repeated deliveries of popular contents substantially contribute to
the surge of multimedia traffic.
For example, there are 8 Billion video requests in Facebook each day,
83$\%$ of which are represented by the top 1$\%$ of the most popular video contents.
To circumvent this retransmission problem, wireless edge caching is advocated
as a powerful technique,
since caching the popular contents can satisfy the majority of video
requests \cite{Wang2020Content}.
\begin{figure*}[t]
\centering{}\includegraphics[scale=0.25]{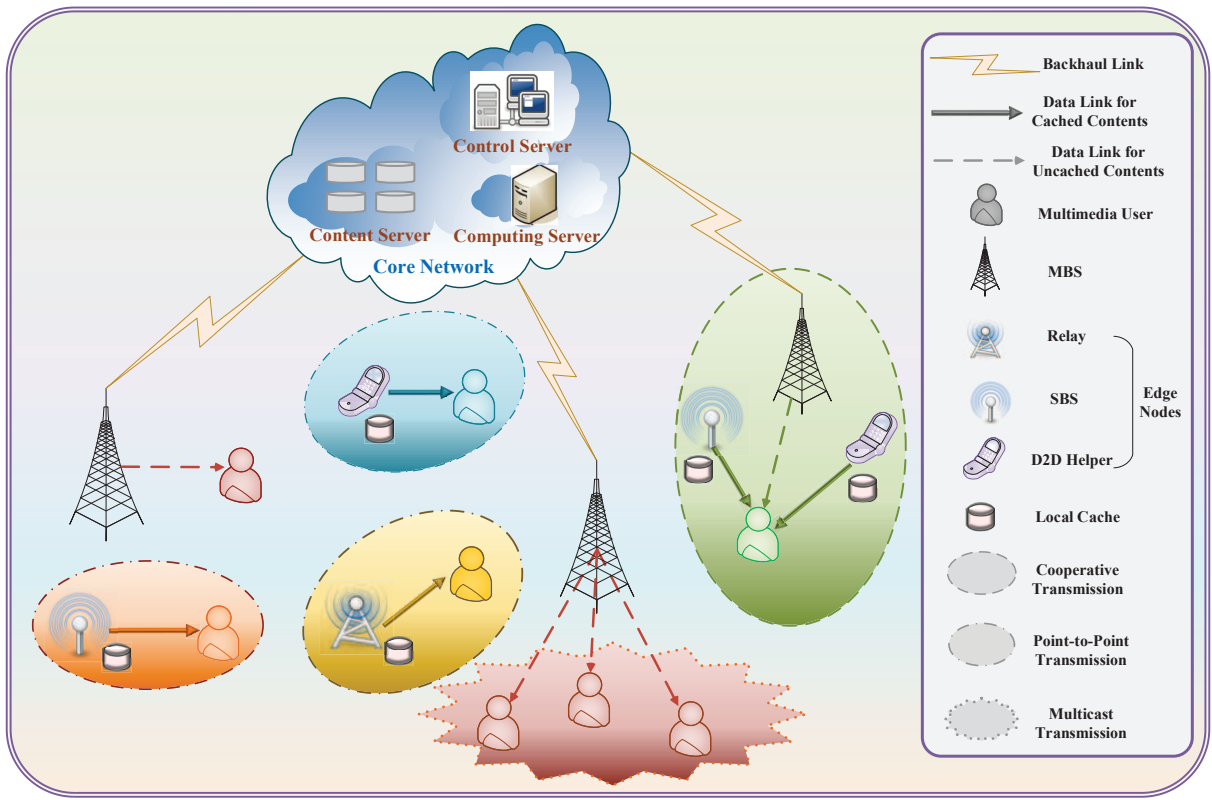}
\caption{An example of wireless edge caching and content transmission scenario,
where the edge nodes roaming in the collaborative area can transmit desired files to users simultaneously.
Multicast transmissions are also adopted to save the system bandwidth and reduce the power consumption.}
\label{System_Model}
\end{figure*}
With the advent of wireless edge caching,
multimedia contents can be pre-fetched and placed in the edge nodes \cite{Wu2019Joint},
such as small-cell base stations (SBSs), device-to-device (D2D) helpers and relays,
which have finite cache capacities.
In general, content placement for wireless caching takes into account
the long-term system states to optimally exploit the finite caching resources
to provide the best multimedia services for end users.
With the aid of edge caching, backhaul congestion can be significantly alleviated,
and the users are able to enjoy flawless multimedia services at a reduced service delay.

An wireless edge caching and content transmission scenario is portrayed in Fig. 1,
where each wireless edge node has finite caching capacity
and the contents to be requested can be pre-stored in the local caches of the nodes.
The macro-cell base stations (MBSs) can provide the users with uncached contents, which are delivered
from the content server through their connected backhauls.
In the collaborative area of Fig. 1,
the nodes storing the requested contents can cooperatively transmit the desired files to the users.
Furthermore, multicast transmissions are also allowed,
where the serving nodes can deliver the requested video contents
to a group of users who have the same watching interests.

In contrast to wired caching,
wireless edge caching has its unique characteristics,
which are summarized as follows.
\begin{itemize}
\item The wireless edge nodes are often power-limited.
      In contrast to the hosts and servers in wired networks,
      the edge nodes, especially the D2D helpers,
      may not have reliable power supply.
      Hence, the power control is a crucial issue
      for wireless caching.
\item In contrast to wired caching networks,
      the network topology of wireless networks is
      generically time-variant due to the mobility of end users.
      To this end, load balancing is required
      between different parts of the wireless edge networks.
\item The hostile wireless channels tend to have a more
grave effect on edge caching and content transmission,
since wireless channels suffer from small-scale and large-scale fading
as well as co-channel interference.
As a result,
the network performance quantified in terms of the transmission delay,
transmit power consumption and energy efficiency (EE)
will be significantly degraded.
Hence, when designing edge caching schemes,
the contents are less likely to be placed at the local caches of
the edge nodes that cause severer fading and interference.
\end{itemize}
Taking the above characteristics into account,
wired caching schemes cannot be directly extended to wireless caching.
The content placement and delivery policies should be specifically designed,
so that the finite cache sizes of edge nodes can be efficiently exploited
in the face of limited power resource,
dynamic network topology and complex channel conditions.

We also have to pay attention to the different viewing quality requirements of end users.
For supporting diverse requirements,
scalable video coding (SVC) can be employed for efficient edge caching design.
The video files encoded by SVC are called scalable videos.
Developed from the advanced video coding (AVC) \cite{Schwarz2007Overview},
SVC can divide a video file into multiple layers.
The layer containing the most fundamental bit stream of the video is termed as the base layer (BL),
because without this layer the so-called enhancement layers (ELs) cannot be decoded for
enhancing the video quality \cite{Boyce2016SHVC}.
The users receiving the BL can only experience standard definition video (SDV),
while those acquiring successive ELs can enjoy high definition video (HDV) \cite{Ostovari2015Scalable}.
SVC provides temporal, spatial and quality scalabilities.
In this work, we rely on SVC's quality scalability
to offer different viewing qualities to the end users.
Compared to the non-scalable H.264/AVC,
the SVC configured for quality scalability will however require extra bits
to provide the same fidelity.
If an optimized encoder control is introduced,
the extra bit rate can be lower than 10\% \cite{Schwarz2007Overview}.
As a compressing benefit,
SVC can cut down the end-to-end latency, boost the fault tolerance
and much more significantly reduce the backhaul pressure.
If the effect of packet loss is considered,
the quality degradation of H.264/AVC will be much more severe than that of SVC.
By integrating SVC with edge caching,
an additional degree-of-freedom can be obtained.
This method does not transmit unnecessary layers,
thus decreasing the power consumption and reducing the content delivery
as well as playout delay experienced by end users.
Some research efforts have been devoted to SVC-based edge caching,
and the related caching schemes under different wireless scenarios are surveyed in Fig. 2.

Although there are benefits in combining SVC with wireless caching,
numerous design challenges also require further consideration.
Firstly, by intrinsically amalgamating SVC with wireless caching,
more SVC-based video layers need to be cached,
which makes the caching schemes more complex.
Moreover, there are multiple viewing quality levels,
and the viewing quality preference has to be modeled or predicted.
Therefore, it is quite challenging to design SVC-based edge caching scheme.
\begin{figure*}[t]
\centering{}\includegraphics[scale=0.99]{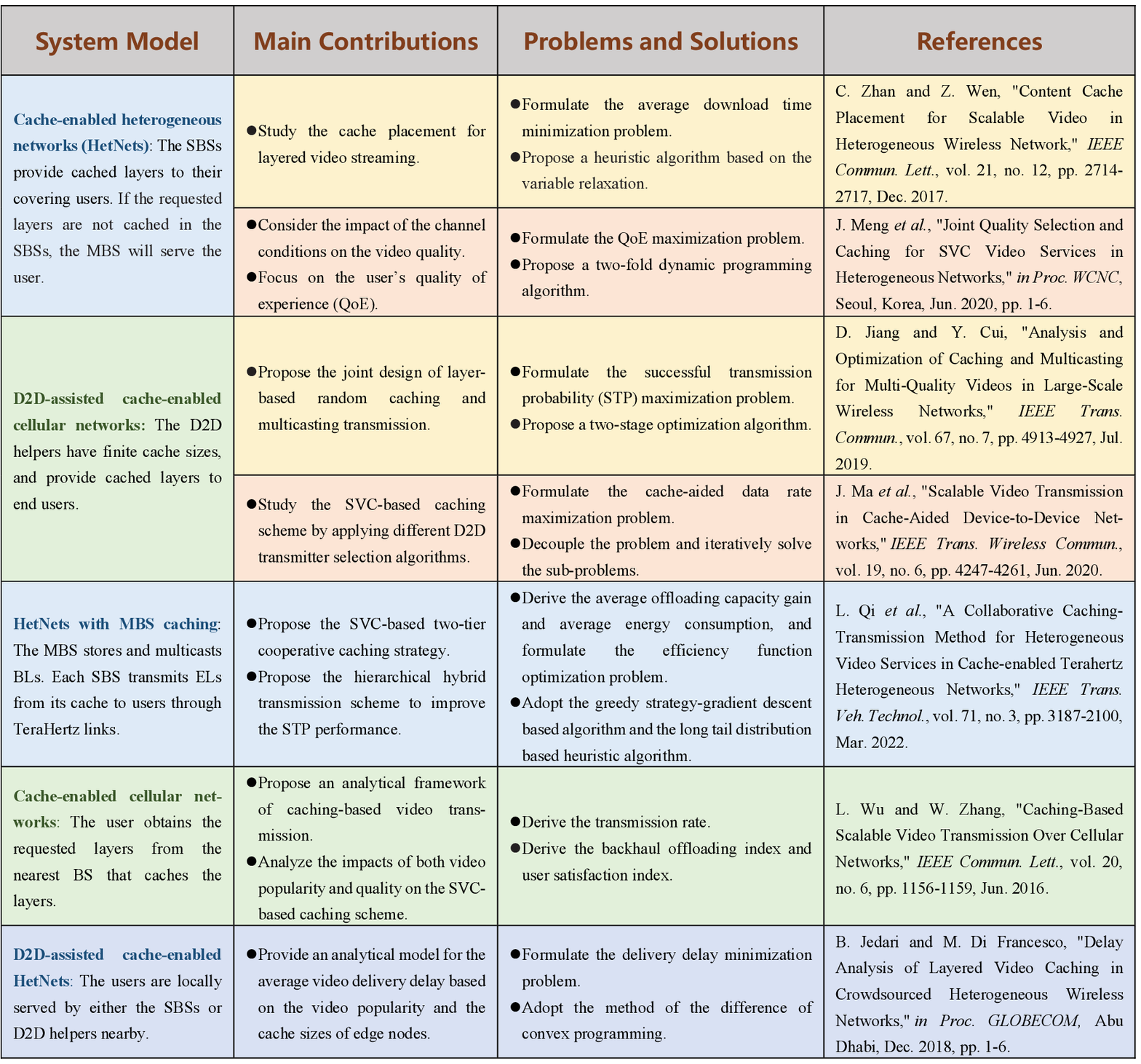}
\caption{The overview of applying SVC into wireless edge caching.}
\label{System_Model}
\end{figure*}

In this light, we investigate the edge caching of scalable videos in wireless networks,
where users can enjoy multi-quality video services based on their preference.
The main contributions of this paper are summarized in the following.
\begin{itemize}
\item We discuss two key issues in SVC-based wireless edge caching,
      namely, the content popularity and quality preference.
      We also highlight the methods of dealing with these two important issues,
      and analyze the differences between caching conventional videos and scalable ones.
\item We reveal the types of cached contents,
      portray the most attractive transmission schemes and summarize the caching decisions
      required for scalable videos.
      The new SVC layer placement strategy is also specified in detail,
      with the aim to provide scalable videos having different viewing qualities for end users.
\item Relying on the proposed SVC-based random caching scheme,
      a case study is conceived for quantifying the transmission delay,
      which is one of the most important performance metrics in wireless edge caching.
      Simulation results confirm that the SVC-based layer placement scheme
      is capable of significantly reducing the transmission delay,
      as compared to the conventional caching scheme operating without SVC.
      We also show that the delay performance of the proposed random caching design
      exceeds that of the benchmark scheme, which only focuses on providing the local caching gain.
\end{itemize}
\section{Key Issues for Wireless Edge Caching}
In SVC-based wireless edge caching,
the video popularity and quality preference are key issues
in determining the layer placement and delivery schemes.
\subsection{Video Popularity Distribution}
The video popularity typically has a strong impact on the content placement,
since popular contents are likely to be cached.
Most wireless cache placement researches rely on the Zipf distribution
to describe the video popularity \cite{breslau1999web}.
In Zipf distribution, the request concentration is reflected by the skewness parameter,
where a higher skewness represents more concentrated video requests.
The Zipf distribution works well for web services in the context of wired caching;
however, it is less applicable for video requests appearing in wireless cellular networks.
Considering this fact, the Mandelbrot-Zipf (MZ) law has been developed
for capturing cellular video requests \cite{Lee2019Throughput}.
Let us assume that there are $F$ video files in the content library,
and the files are ranked in descending order of popularity.
Following the MZ distribution, the request probability of the $f$-th video file is given by
$P_{f}=\frac{(f+q)^{-\alpha}}{\sum_{n=1}^{F}(n+q)^{-\alpha}},\ f=1,2,...,F,$
where $\alpha$ is the skewness parameter and $q$ is the plateau factor.
With the increase of $q$, the difference of the request probabilities decreases among the popular files.
Also note that when the plateau factor is zero,
the MZ distribution becomes the conventional Zipf distribution.
The parameters of the distributions are time-varying,
hence they should be specified empirically by applying curve-fitting methods based on real-world datasets.

Given a specific experimental popularity distribution,
we can readily determine the important parameters for accurately characterizing the video request popularity.
In other cases, however, the existing distributions cannot accurately reflect the video popularity.
In this case, with multiple user requests collected from wireless edge nodes or content server,
the video popularity can be estimated by machine learning algorithms \cite{Chen2018Caching}.
Based on the predicted popularity distribution,
low-complexity caching placement strategies,
such as the conventional most popular content placement (MPCP),
are available at the edge nodes.
\subsection{Heterogeneous Quality Preference}
In the presence of diverse viewing requirements,
the quality preferences of end users deserve careful consideration.
The quality preference is typically predicted either by
using curve-fitting methods or machine learning algorithms.
To predict the quality preference, numerous factors have to be jointly taken into account,
such as the types of video files and users as well as the screen sizes.

The specific type of the video files is one of the most primary factors that affects the quality preference.
A low viewing quality may be adequate for news,
but high perceptual experiences are preferred for entertainment programs,
such as movies, TV series and variety shows.
In this case, more layers of the SVC-based video files are requested
to satisfy the viewing quality requirements,
which increases the request probability of the ELs belonging to these video files.

The quality preference also depends on the type of users.
Typically, users having good channel conditions are likely to request HDVs
for superb viewing experiences, and more ELs are requested and delivered.
Often, superior perceptual qualities are delivered to the VIP users, e.g., those who pay premiums.
Owing to the fact revealed above,
the quality preference differs between different classes of users.

The screen size is the third factor.
The devices having large screen sizes usually require HDVs,
since they are sensitive to grainy viewing qualities.
By contrast, devices with smaller screen sizes may tolerate SDVs.
\section{Content Delivery Process for Scalable Videos}
Given the layered structure of scalable videos,
the content delivery processes are different
from those of their traditional counterparts.
Before designing the content delivery schemes,
a discussion about the types of cached contents is provided.
\subsection{Types of Cached Contents for Scalable Videos}
By applying the SVC technique, each video file is encoded into multiple layers.
The SVC-based layers can be locally cached,
and the edge nodes can store the BL and multiple ELs for each scalable video,
each of which is regarded as a caching unit.
After end users receive a sufficient number of layers to satisfy their viewing requirements,
the SVC decoding can be performed to recover the requested videos having the preferred quality levels.

As a design alternative,
a set of successive layers can also be viewed as a caching unit,
which we refer to as a super layer (SL) \cite{Guo2018Multi}.
Each SL consists of a single BL and a number of successive ELs,
where the specific number of ELs is determined by the particular viewing quality requirements of users.
When the SL is transmitted to the user,
it is capable of providing the requested video file
at any of the preferred viewing quality by
delivering the BL and any number of ELs.
When SLs are cached and transmitted to users,
the encoding and decoding processes are completed at the content server in the core network in advance.
More specifically, the original video is firstly partitioned into multiple layers by SVC,
and the successive SVC-based layers are then
decoded into the videos having different viewing qualities, i.e., SLs,  by SVC decoding.
As a benefit, if SLs are provided,
seamless video services can be provided,
and the decoding process is conveniently avoided at the receivers' side.

Naturally, there are a number of trade-offs.
Firstly, it is worth pointing out that
caching the individual SVC-based layers can enhance the content diversity,
as compared to storing the SLs,
since the size of the SL is typically larger than that of its constituent BL and ELs.
This indicates that, when the edge nodes are equipped with smaller cache sizes,
it is a wise choice to store individual BLs and ELs.
Moreover, when the computational capability of receivers is limited,
the edge nodes prefer to cache SLs instead.
\begin{figure*}[t]
\centering{}\includegraphics[scale=0.45]{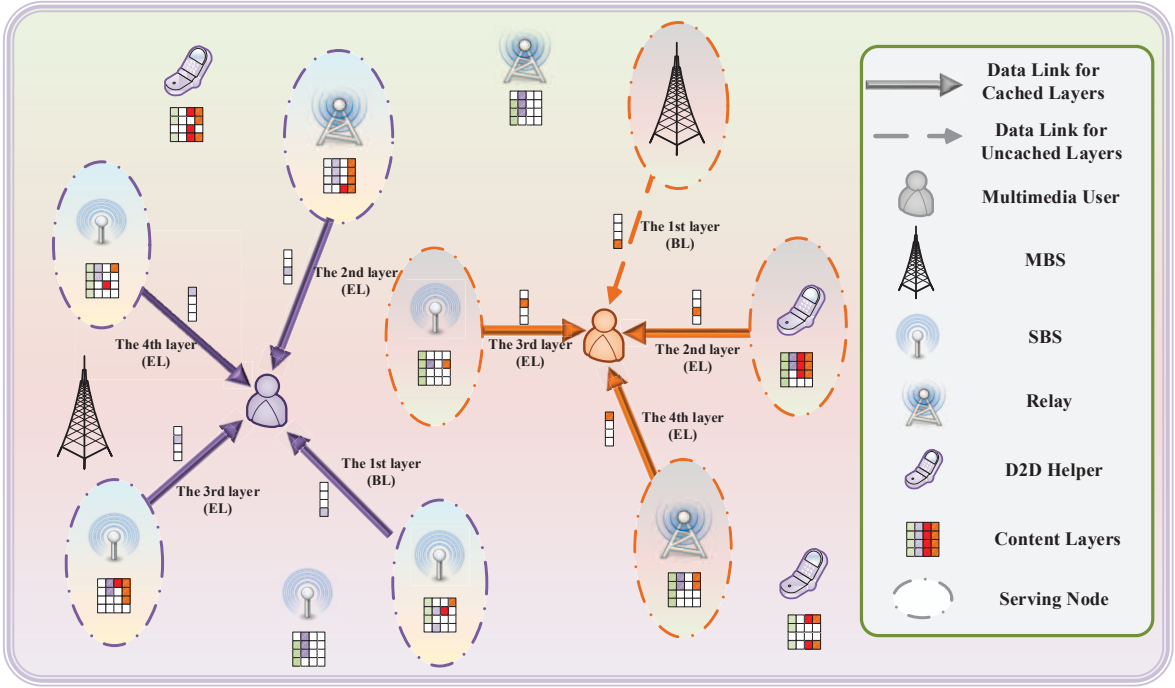}
\caption{An example of the ILT scheme,
where each video is partitioned into four layers
and the users request the video files colored in purple and orange.
The SBSs caching the requested layers transmit desired layers to users simultaneously,
and the un-cached layers are retrieved and transmitted by the MBS.}
\label{System_Model}
\end{figure*}
\subsection{Content Delivery Designs}
Given the different kinds of cached contents,
the transmission policies can be accordingly designed.

{\bf{Scenario 1}}: If multiple individual BLs and ELs are cached
in the local caches of edge nodes,
the nodes can individually transmit the requested layers to the user.
This design is referred to as individual layer based transmission (ILT) in this article.
Typically, the serving nodes share the same frequency and time resources.
Hence, they can be tightly coordinated by the MBS covering these nodes.
At the beginning of each time slot,
the nodes caching the desired layers are synchronized,
and then the requested layers are transmitted to the user.
When the user receives multiple super-imposed content layers,
successive interference cancellation (SIC)
is required for separating the layers \cite{Xu2017Modeling}.
Then the user can decode the desired video file at preferable
viewing quality by performing SVC decoding.
An example of this kind of content transmission design is shown in Fig. 3.

{\bf{Scenario 2}}: In another case where the SLs are locally stored,
the cached SLs can be delivered by the serving nodes.
This content delivery strategy is referred to as SL based transmission (SLT).
The nodes having the requested SLs in their caches can serve the end users,
and the cooperative transmission and multicast transmission are also allowed
to deliver the requested SLs.
The user, who receives the SL,
can directly view the requested multimedia video without SIC and SVC decoding.
As a result, the playback performance can be substantially improved.
This kind of transmission design is portrayed in Fig. 4.

\begin{figure*}[t]
\centering{}\includegraphics[scale=0.42]{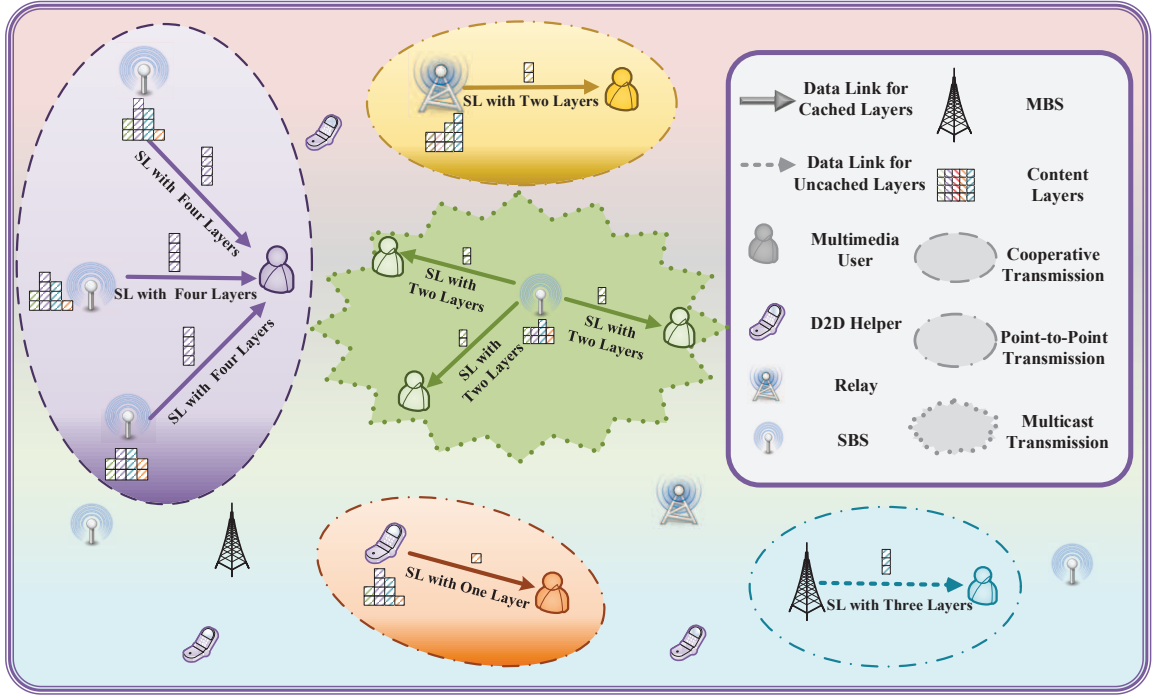}
\caption{An illustration of the SLT scheme,
in which the cooperative transmission, multicast transmission and point-to-point transmission are allowed.}
\label{System_Model}
\end{figure*}

Both Scenarios 1 and 2 have their pros and cons.
Although the ILT scheme requires a complex receiver for SIC and SVC decoding,
it allows for the cooperation of multiple serving nodes,
hence avoiding a single-point failure.
Compared to the ILT, the SLT scheme simplifies the receiver design,
which therefore reduces the playout delay.
However, when employing the SLT scheme, if the serving node fails to provide the requested content,
the user would require retransmissions and suffer from longer delays in retrieving the requested content.
This is different from using the ILT scheme, which
only has to deliver the missing BL or ELs.
In practical implementations,
the specific choice of the type of cached contents primarily depends on
the network topology, the receiver design and the playout requirements of end users.
\section{Caching Decisions for Scalable Videos}
In this section, we first review the existing wireless caching schemes conceived for salable videos,
and then present a new caching strategy.
The design criteria of the caching decisions are also summarized.
\subsection{Caching Schemes for Scalable Videos}
We first discuss the caching schemes suitable for scalable videos.
Typically, depending on the types of caching decisions,
the caching schemes can be classified into two categories.
In the first category, the caching decisions are continuous variables, ranging from 0 to 1.
There are two popular caching schemes in this category,
namely, fractional caching \cite{Xu2017Modeling} and random caching \cite{Zhang2018Energy}.
The details of these two schemes are listed as follows.
\begin{itemize}
\item {\bf{Fractional Caching}}. Under this caching scheme,
      each single layer or SL is partly cached in the local caches of edge nodes.
      The caching fraction for each layer is determined later.
      The uncached parts of the requested contents have to be delivered from the core network via the backhauls.
      Typically, fractional caching is suitable for layers with large sizes
       to improve the layer diversity and content hit probability.
\item {\bf{Random Caching (also termed as probabilistic caching)}}. Based on this caching strategy,
      the edge nodes can probabilistically store the single layers or SLs
      based on the caching probability distributions.
      The layer having a higher caching probability is more likely to be locally cached.
      This caching scheme is capable of operating in the face of user mobility \cite{Yao2019On}.
\end{itemize}

Furthermore, there is also a new type of layer placement for scalable videos,
where the caching decisions are binary variables,
i.e., either 0 or 1.
Specifically, if the caching indicator is 1,
the corresponding layer will be locally cached.
Otherwise, the layer cannot be found in the local caches.
In most cases, the optimization problems associated with binary variables are NP-hard,
and cannot be efficiently solved in polynomial time.
To this end, it is important to design efficient layer placement schemes
operating at a reduced computational complexity.
\subsection{Caching Criteria for Scalable Videos}
Under the above caching schemes,
the remaining problem is how to determine the caching decisions for the edge nodes.
For different load balancing states, user requirements and power consumption levels,
different caching criteria have to be considered.
In the literature, the following caching criteria have been studied:
\begin{itemize}
\item {\bf{Average download time}} \cite{Zhan2017Content}.
      When SVC-based caching schemes are adopted at the edge nodes,
      unnecessary ELs do not have to be delivered,
      thus reducing the content download delay.
      If ILT is adopted, the download delay depends on the maximum time of downloading each requested layer;
      otherwise, the download delay is determined by the total time for downloading the requested SLs.
\item {\bf{Transmit power consumption}} \cite{Guo2018Multi}.
      In contrast to the video files without SVC processing,
      caching and transmitting SVC-based layers
      can decrease the power consumption of edge nodes.
      Specifically, multicast transmission can be adopted for
      further reducing the total power consumption \cite{Guo2018Multi}.
\item {\bf{EE}} \cite{Zhang2018Energy}.
      As revealed in \cite{Zhang2018Energy},
      there is a trade-off between the data rates of end users and the total power consumption,
      and the power savings should not degrade the users' QoS.
     In this sense, the EE has to be carefully considered
      when providing scalable video services.
      It is worth noting that the cooperation among multiple serving nodes
      should also be taken into account,
      when maximizing the EE in the scenario
      where individual layers are cached and transmitted simultaneously.
      Moreover, when establishing the total power consumption model,
      the consumption arising from content caching and backhaul deliveries
      should also be accurately modeled.
\item {\bf{Content transmission delay}} \cite{Li2018Learning}.
      In modern cellular networks,
      the finite capacity of the backhaul may lead to an adverse effect on the content transmission delay.
      Severe backhaul congestion usually results in an unacceptable latency.
      When scalable videos are cached at the wireless edge,
      users can retrieve part of the requested contents without backhaul involvement,
      hence the transmission delay caused by backhaul-based retrieval can be avoided.
      Moreover, since the number of transmitted layers is tailored for each user,
      superfluous layers are not delivered,
      which further reduces the transmission delay.
\end{itemize}
\section{Case Study of SVC-based Wireless Caching}
In this section,
we present a case study to validate the benefits of caching and transmitting scalable videos.
\subsection{System Model}
{\bf{Network Model:}} A three-tier heterogeneous network,
consisting of a D2D tier, an SBS tier and an MBS tier, is observed.
The three-tier network model considered is practical,
and it may be readily extended to multi-tier models or simplified to two-tier models.
For the different number of tiers and for diverse node deployments,
the content caching and transmission designs have to be appropriately adjusted,
based on similar steps to those advocated in this paper.
The locations of edge caching nodes follow the classic Poisson point processes (PPPs)
with different distribution densities.
The density of D2D helpers is the highest among the three types of nodes,
while the MBSs have the lowest distribution density.
The fading and interference inflicted by wireless channels are modeled as follows.
The small-scale fading follows the Rayleigh distribution with zero mean and unit variance,
i.e., $\mathcal{CN}(0,1)$;
the pathloss is described by $d^{-\alpha}$,
where $d$ denotes the distance between the node and the user
and $\alpha$ is the pathloss exponent;
finally, the co-channel interference experiencing both small-scale and large-scale fading
is also included in the denominator of the signal-to-interference-plus-noise ratio (SINR)
experienced by the user.

{\bf{Transmission and Caching Policies:}}
The D2D helpers and SBSs are equipped with finite caching storage.
They can provide cached layers to nearby users.
The D2D helpers and SBSs that are capable of providing cached layers
are distributed in the serving areas having radii $r_{d}$ and $r_s$ around the requesting user.
If the requested layers can be found in the nearby D2D helpers,
the nearest D2D helper in the vicinity of the user and caching the requested layers
is selected as the serving node.
If the layers cannot be found in the caches of the D2D helpers,
the user will resort to fetching them from the SBSs.
When edge nodes provide the users with requested layers,
the transmission delay is caused by the downlink transmission from the nodes to the users.
However, if some of the requested layers are not cached locally,
the nearest MBS around the user will provide the missing layers
from the content server via the backhaul.
Apart from the delay of downlink transmission from the MBS to users,
the delay contributed by backhaul retrievals is also non-negligible in this case.
Regarding the caching scheme, the random caching strategy is adopted,
where SVC-based layers are probabilistically cached in the edge nodes.

{\bf{Problem Formulation and Solutions:}}
Under the proposed transmission and caching policies,
the optimization problem of transmission delay minimization is formulated,
with the constraints of finite cache sizes of each D2D helper and SBS.
The optimization problem is solved
by the standard gradient projection method,
which is eminently efficient in solving our optimization problem
having a differentiable objective function over a convex variable set \cite{Zhang2018Energy}.
Then, the random caching probability of each layer is found.
In the proposed algorithm,
the partial derivatives of the transmission delay
with respect to caching probabilities have to be calculated for each iteration.
The computational cost of the partial derivative calculations
dominates the complexity of the algorithm,
and the number of calculations is proportional to $F\cdot L$,
where $F$ and $L$ are the total numbers of video files and content layers, respectively.
Thus, when employing the ${\cal O}$ function for our complexity analysis,
the overall overhead of the proposed algorithm is ${\cal O}(F\cdot L)$.
\subsection{Simulation Results}
\begin{figure}[t]
\centering{}\includegraphics[scale=0.65]{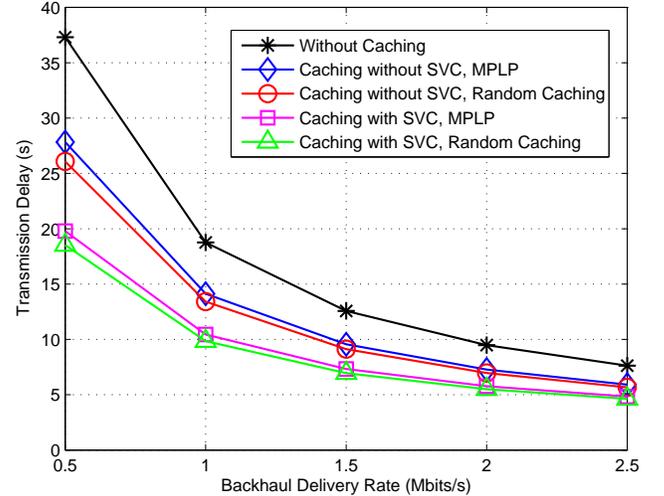}
\caption{The transmission delay versus the backhaul delivery rates.}
\label{System_Model}
\end{figure}
In the simulations, the size of each video without SVC processing is 50 Mbits.
Inspired by \cite{Schwarz2007Overview},
we set the size of each SVC-based video file to be 10\% larger than that of the file without SVC.
For each scalable video, it is partitioned into 8 layers.
The radii of the serving areas for the D2D helpers and SBSs are $10$ m and $30$ m, respectively.
The path-loss exponent is 4 for D2D helpers, SBSs and MBSs.
The cache sizes of each D2D helper and SBS are 200 Mbits and 500 Mbits.
We adopt the most popular layer placement (MPLP) as our benchmark scheme,
where the SVC-based layers from the most popular videos
fully occupy the local caches of D2D helpers and SBSs.

In Fig. 4, we present the relationship between the content transmission delay and backhaul delivery rate.
Clearly, a high backhaul delivery rate will reduce the transmission delay.
In the benchmark schemes, the scheme without caching yields the worst delay,
since all requested layers have to be retrieved by capacity-limited backhauls.
We can also find that by providing SVC-based video services,
the transmission delay can be significantly decreased.
This is because un-requested layers are not delivered to end users.
Compared to the MPLP scheme, random caching scheme can indeed reduce the transmission delay.
The MPLP only focuses on providing the local caching gain that each individual node can provide,
while the random caching can exploit the aggregated cache sizes of all caching nodes.
As a result, random caching significantly reduces the transmission delay.

\begin{figure}[t]
\centering{}\includegraphics[scale=0.65]{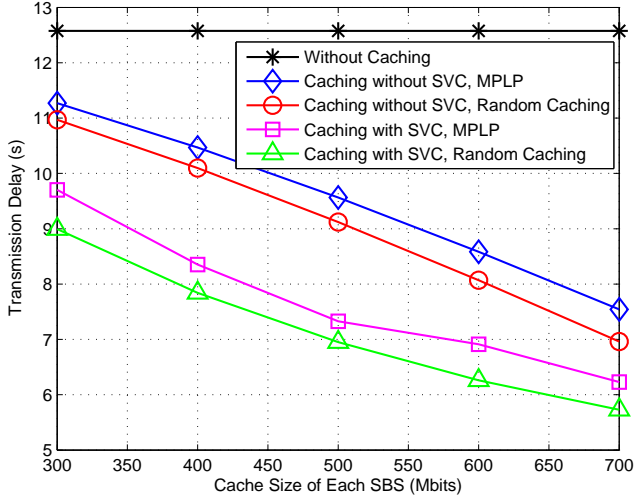}
\caption{The transmission delay versus the cache sizes of SBSs.}
\label{System_Model}
\end{figure}

The relationship between the content transmission delay and the cache size of each SBS
is characterized in Fig. 5.
Observe that the transmission delay decreases as the cache size of each SBS grows,
since more cached layers are provided for end users
and repeated backhaul deliveries are substantially reduced.
Likewise, the delay of random caching is lower than that of MPLP,
and the SVC-aided caching scheme imposes lower transmission delay than that without SVC.
\section{Conclusions and the Future Directions}
In this article, wireless edge caching designed for scalable videos was studied.
We discussed the issues of content popularity as well as quality preference,
and collated the types of cached contents and their corresponding transmission schemes.
We elaborated on the caching options in support of scalable video services
and the performance metrics used for making caching decisions.
Finally, a case study was presented to validate the benefits
of caching and transmitting SVC-based layers.
From our simulation results, we concluded that
caching and transmitting scalable videos is capable of reducing the transmission delay.
Additionally, it can be observed that the transmission delay of the proposed
random caching scheme is lower than that of the caching scheme
which only focuses on providing the local caching gain.

As a promising technique to
reduce repeated content deliveries and relive the backhaul pressure
in future wireless networks,
there still remain some crucial challenges to be solved in the field of edge caching,
listed as follows.
\begin{itemize}
\item The transmission reliability of the BL is not well documented in the literature,
even though the BL is the most essential part of scalable video streamings.
When offering scalable video services to end users,
how to guarantee the BL is reliably transmitted and decoded remains a future challenge.
\item The trade-off between encoding/decoding complexity and the performance improvement attained
needs a balanced consideration.
With more SVC-based layers,
the system performance will be improved,
since more layers can be flexibly cached and transmitted.
However, the encoding/decoding complexity escalates with
the number of divided layers.
Hence, a compelling trade-off between the encoding/decoding complexity
and the performance improvement is desired.
\item When designing caching schemes,
having efficient power control is another crucial issue that deserves further investigations.
Firstly, the edge nodes providing the scalable video services
are often power-limited, especially the D2D helpers,
and hence the power control plays a prominent role.
Power control actions typically have a shorter time scale than content caching,
which imposes a challenge when jointly designing the power control and content caching schemes.
Additionally, for SVC-based edge caching,
the power consumption, video quality and transmission delay are important aspects
that have to be jointly considered.
These metrics are often conflicting and coupled with each other.
To strike an attractive trade-off among them,
it is of prime concern to formulate a multi-objective optimization problem,
which may be solved by the sophisticated multi-objective evolutionary algorithm.
Alternatively, it may be transformed into a single-objective optimization problem
associated with $\varepsilon$ constraints.
If there are a high number of parameters,
machine learning (ML) may be used for solving large-scale Pareto optimization problems.
With these effective algorithms,
all Pareto optimal points of the Pareto front can be found.
\item  The security and privacy of end users and D2D helpers may not be guaranteed.
On one hand, when using machine learning approaches for
predicting the video popularity and quality preference,
a large amount of information will be collected
by the edge nodes or content server.
The collected information is personal and private, and may be wiretapped by the eavesdroppers.
On the other hand, when D2D helpers assist in caching and transmitting SVC-based layers,
the personal information of the helpers may be exposed to end users,
which also causes rather severe security and privacy problems.
Instead of our helper selection scheme, helpers may also be selected for improving the information security and minimizing the eavesdropping probability.
Furthermore, the secrecy rate,
together with the power consumption, the video quality and the transmission delay,
could also be incorporated into the multi-objective optimization problem.
\end{itemize}
These issues deserve research attention and will guide our future directions.
\section*{Acknowledgement}
X. Zhang and Y. Ren would like to acknowledge the financial support of the National Natural Science Foundation
of China under Grants 62101442, 61801382, 62071377 and 62001264,
the Natural Science Foundation of Shaanxi Province under Grant 2021JQ-705,
and the Young Talent fund of University Association for Science
and Technology in Shaanxi, China under Grant 20210115.

L. Hanzo would like to acknowledge the financial support of the Engineering and Physical Sciences Research Council projects EP/W016605/1 and EP/P003990/1 (COALESCE) as well as of the European Research Council's Advanced Fellow Grant QuantCom (Grant No. 789028).

\bibliographystyle{IEEEtran}

\section*{Biographies}
\noindent {\small {}Xuewei Zhang (zhangxw@xupt.edu.cn)
received the Ph.D. degree in information and communication engineering
from the Beijing University of Posts and Telecommunications (BUPT), Beijing, China, in 2020.
She is currently an Associate Professor with the School of Communications and Information Engineering,
Xi'an University of Posts and Telecommunications (XUPT), Xi'an, China.
Her research interests include wireless edge caching, resource allocation and heterogeneous networking.}
\\

\noindent {\small {}Yuan Ren
received the B.Eng. degree in information engineering
and the Ph.D. degree in signal and information processing
from the Beijing University of Posts and Telecommunications (BUPT),
Beijing, China, in 2010 and 2017, respectively.
He is currently an Associate Professor with the School of Communications and Information Engineering,
Xi'an University of Posts and Telecommunications (XUPT), Xi'an, China.
His current research interests include green communications,
wireless caching and cooperative communications.}
\\

\noindent {\small {}Tiejun Lv
received the M.S. and Ph.D. degrees in electronic engineering from
the University of Electronic Science and Technology of China (UESTC),
Chengdu, China, in 1997 and 2000, respectively.
He is the author of 3 books, more than 100 published IEEE journal papers
and 200 conference papers on the physical layer of wireless mobile communications.
His current research interests include signal processing, communications theory and networking.}
\\

\noindent {\small {}Hanzo Lajos
received his doctorate in 1983. He holds an
honorary doctorate by the Technical University of Budapest
(2009) and by the University of Edinburgh (2015). He is a
member of the Hungarian Academy of Sciences and a former
EIC of the IEEE Press. He is a Governor of both IEEE
ComSoc and VTS.}
\end{document}